**Model Validation and Selection in Metabolic Flux Analysis and Flux Balance Analysis**

Joshua A.M. Kaste[1,2,*], Yair Shachar-Hill[2*]

[1]Department of Biochemistry and Molecular Biology, Michigan State University, 603 Wilson Rd, East Lansing, MI 48823

[2]Department of Plant Biology, Michigan State University, 612 Wilson Rd, East Lansing, MI 48824

*corresponding authors

**Abstract**

13C-Metabolic Flux Analysis (13C-MFA) and Flux Balance Analysis (FBA) are widely used to investigate the operation of biochemical networks in both biological and biotechnological research. Both of these methods use metabolic reaction network models of metabolism operating at steady state, so that reaction rates (fluxes) and the levels of metabolic intermediates are constrained to be invariant. They provide estimated (MFA) or predicted (FBA) values of the fluxes through the network *in vivo*, which cannot be measured directly. A number of approaches have been taken to test the reliability of estimates and predictions from constraint-based methods and to decide on and/or discriminate between alternative model architectures. Despite advances in other areas of the statistical evaluation of metabolic models, validation and model selection methods have been underappreciated and underexplored. We review the history and state-of-the-art in constraint-based metabolic model validation and model selection. Applications and limitations of the $\chi^2$-test of goodness-of-fit, the most widely used quantitative validation and selection approach in 13C-MFA, are discussed, and complementary and alternative forms of validation and selection are proposed. A combined model validation and selection framework for 13C-MFA incorporating metabolite pool size information that leverages new developments in the field is presented and advocated for. Finally, we discuss how the adoption of robust validation and selection procedures can enhance confidence in constraint-based modeling as a whole and ultimately facilitate more widespread use of FBA in biotechnology in particular.

**Introduction**

The set of biochemical reaction rates in the metabolic network of a living system (its flux map) represents an integrated functional phenotype that emerges from multiple layers of biological organization and regulation, including the genome, transcriptome, and proteome [1]. The study of metabolic fluxes is therefore important for systems biology, rational metabolic engineering, and synthetic biology. A grand challenge of systems biology is building an integrated mechanistic understanding of the operation of living organisms across these levels of regulation [2] – an understanding that goes beyond statistical or correlative descriptions, however useful these can be. Meeting this challenge requires fluxes to be accurately predicted from network structure using explicit rules or hypotheses and reliably estimated using experimental data. Fluxes are also critical to many biotechnological and metabolic engineering applications. Examples such as the development of lysine hyper-producing strains of *Corynebacterium glutamicum* [3–5] and the rewiring of *E. coli*'s metabolism to make it grow chemoautotrophically [6] attest to the usefulness of these techniques. As the scale and complexity of integrative systems biology and biological engineering efforts increase, so too will the need for reliable and robust estimates of fluxes.

I*n vivo* fluxes cannot be directly measured, necessitating modeling approaches to estimate or predict them. The most commonly used approaches for metabolic modeling are the constraint-based modeling frameworks of 13C-Metabolic Flux Analysis (13C-MFA) and Flux Balance Analysis (FBA). Both require a metabolic network consisting of metabolites linked by biochemical reactions to be defined using the biochemical literature, knowledge of the enzymes and transporters expressed from the genome and physico-chemical rules. In 13C-MFA, atom mappings describing the positions and interconversions of the carbon atoms in reactants and products are also included in the model. These methods assume that the system is at metabolic steady-state, such that the concentrations of all metabolic intermediates and reaction rates are constant [7]. External fluxes, such as the uptake of a substrate or the rate of production of new cells or a product, are also measured and constrained. These assumptions and constraints define a "solution space" containing all flux maps consistent with them but are typically insufficient to pinpoint a unique flux map.

In 13C-MFA, isotopic labeling data is used to identify a particular solution within the solution space. $^{13}$C-labeled substrates are fed to the system under investigation and the endpoint labeling, or time-course labeling in Isotopically Nonstationary Metabolic Flux Analysis (INST-MFA), of metabolites is measured using mass spectrometry and/or NMR techniques [7,8]. Given a metabolic network, a flux map, and information about the labeled substrate fed into the system, the label distribution through all the metabolites in a network can be solved analytically. However, 13C-MFA works backwards from measured label distributions to flux maps by minimizing the residuals between measured and estimated Mass Isotopomer Distribution (MID) values by varying flux and pool size estimates [9]. For INST MFA pool size measurements can also be included in the minimization process.

In FBA, linear optimization is used to identify a flux map (or set of flux maps) from the solution space [10]. This is the map(s) for which the sum of one or more fluxes (the objective function) is maximized or minimized. Objective functions frequently represent measures of efficiency, including the maximization of growth rate or product formation or the minimization of total flux [11]. Such functions may embody hypotheses about what the *in vivo* system has been evolutionarily tuned to optimize, or questions about the operational capacity of that system under particular conditions. In many cases, the constraints – typically on external fluxes – imposed during an FBA optimization result in a set of viable flux maps (a solution space) rather than a single map. In such cases, related techniques, including Flux Variability Analysis [12] and random sampling [13–16] can be used to characterize the set of flux maps consistent with the set constraints. The computational tractability and small amount of experimental data necessary to perform FBA allow the analysis of Genome-Scale Stoichiometric Models (GSSMs). These models incorporate all the reactions believed to occur in an organism based on a combination of genome annotation and manual curation. Additional linear-optimization based methods for solving GSSMs using the FBA framework have been developed and are sometimes used together with FBA. These include Minimization of Metabolic Adjustment (MOMA) [17], and Regulatory On/Off Minimization (ROOM) [18], as well as a host of methods that incorporate omic data into the optimization process (e.g. [19–23]). FBA and its related methods, are sometimes used to analyze models other than true GSSMs, such as "core" models that focus on central metabolic processes that conduct the large majority of flux [24]. When discussing validation, however, the same principles apply to all of these linear optimization methods and across the different model scales. For the sake of simplicity, we will be using "FBA" to refer to this family of methods generally and will refer to the medium- to large-scale models used with these methods as "FBA models."

Progress has been made in improving the statistical rigor and reliability of flux estimates. For example, the development of effective methods for flux uncertainty estimation [25] allows researchers to better quantify confidence in flux predictions and, where appropriate, to gather additional data to better support their conclusions. Related to this are advances in designing and implementing parallel labeling experiments, wherein multiple tracers are employed in parallel labeling experiments and the results are simultaneously fit to generate a single 13C-MFA flux, enable more precise estimation of fluxes than experiments with individual tracers or tracer combinations allow [26–33].

However, other areas of the statistical evaluation of constraint-based modeling studies have received less discussion in the literature. How can MFA and FBA researchers validate the accuracy of their estimates and predictions? These methods also require researchers to make choices about the network structure of the model to be used. This leads to questions of model selection; that is, how do we select the most statistically justified model from among the alternatives? Validation and model selection are key to improving the fidelity of model-derived fluxes to the real *in vivo* ones. Despite substantial development of model selection and validation practices in systems and synthetic biology [34,35], the flux analysis community has comparatively few consistent practices or guidelines. Addressing these topics explicitly is also important for readers of the flux analysis literature to understand the assumptions, tests of validity, and model selection techniques underlying what they are reading.

We review and provide our perspective on these areas and prospects for future development, highlighting: (1) Validation methods applicable to FBA flux maps; (2) approaches for validating 13C-MFA flux maps; and (3) developments and prospects for model selection in 13C-MFA; (4) How validation and model selection practices in 13C-MFA could benefit from a greater emphasis on the isolation of training and validation datasets and; (5) the importance of corroborating flux mapping results using independent modeling and experimental techniques.

## Validation Techniques in FBA and 13C-MFA

FBA and 13C-MFA studies commonly validate the model(s) used, though there is great variation in their nature and extent. We summarize these validation strategies in **Figure 1**.

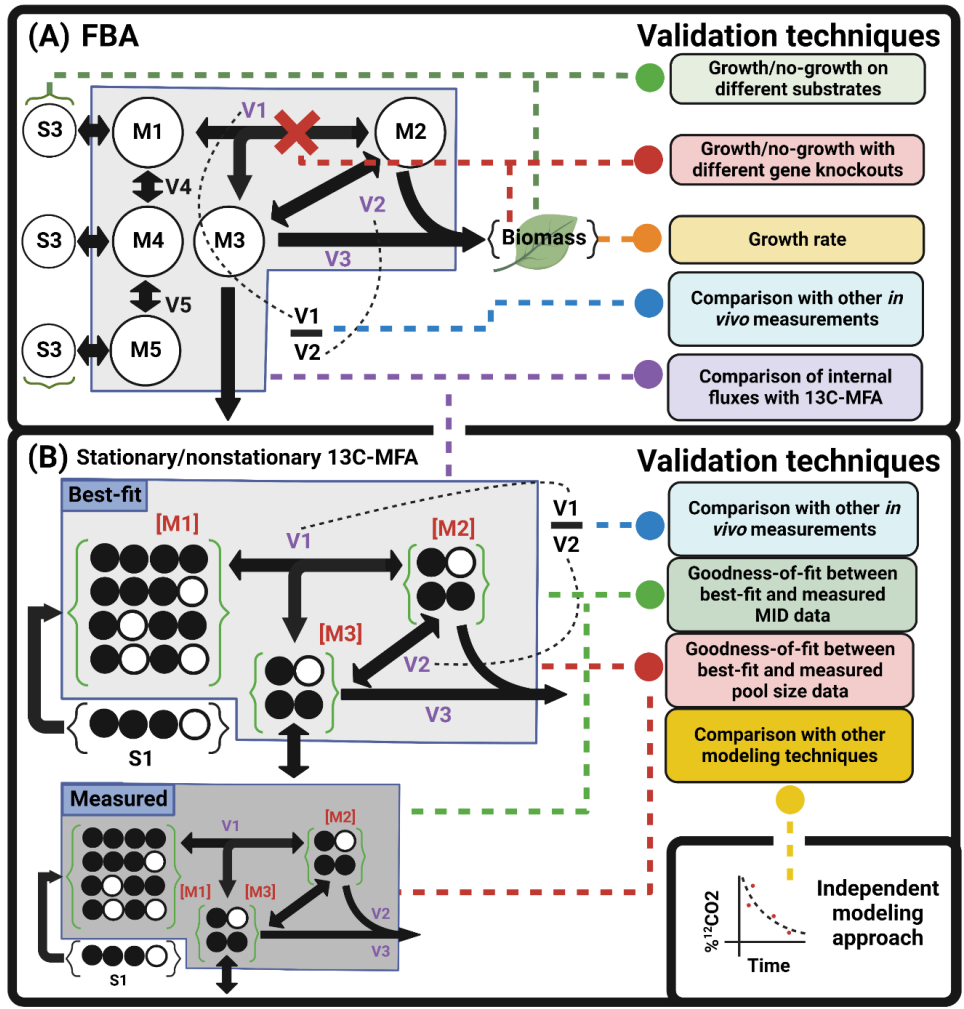

**Figure 1**: *Graphical summary of validation strategies in **(A)** FBA and **(B)** 13C-MFA. Dotted lines connect inputs with the associated validation technique(s). **(A)** FBA predictions can be validated by comparing growth rate or growth/no-growth phenotypes across different substrates, growth conditions, or sets of gene knockouts in silico and in vivo. Values can be calculated from flux maps and compared with experimental measurements. FBA internal flux predictions can be compared with 13C-MFA fluxes. **(B)** Values can be calculated from 13C-MFA flux maps and compared with an independent experimental measurement from the in vivo system. Goodness-of-fit can be assessed between simulated and measured MIDs, and simulated and measured metabolite pool sizes in INST-MFA. Flux maps can be compared with the results of independent modeling exercises. Molecules are schematically shown as connected circles of atomic positions: open circles are unlabeled, filled circles are isotopically labeled. Abbreviations: $M_n$ - metabolites in the metabolic network; $S_n$ – exogenous substrates; $V_i$ – Fluxes; $[M_n]$ – metabolite concentrations.*

*Validation in FBA*

The COnstraint-Based Reconstruction and Analysis (COBRA) framework, implemented in software solutions such as the COBRA Toolbox [36] and *cobrapy* [37] and widely used for FBA studies, features functions and pipelines that can be used to ensure basic functionality of models. Model characteristics evaluated include the inability to generate ATP without an external source of energy and the inability to synthesize biomass without adding substrates not known to be needed. Additionally, the MEMOTE (MEtabolic MOdel TEsts) pipeline contains tests to ensure, for example, that biomass precursors can be successfully synthesized in a model in a variety of growth media [38]. MEMOTE has been used as a way of ensuring appropriate stoichiometry and consistency with accepted format standards in models entered into the BiGG [39] model database. These forms of Quality Control are an important first step in ensuring that models are behaving appropriately and generating useful predictions; however, following these initial checks on functionality, the techniques used to validate actual model predictions, are varied and not standardized. Indeed, even in the BiGG database, which is highly curated and focuses primarily on models of microbial systems, models vary in the type and extent of validation performed. Given the variety of validation procedures that appear in the literature, it is important when using an FBA model to be aware of what specific validations were used, their limitations are, and consequently, what inferences or downstream applications are appropriate (summarized in **Table 1**).

Perhaps the most common validation in FBA is comparison between FBA-predicted and empirically measured rates of growth (e.g. [40–43]) or product formation. One may similarly evaluate growth/no-growth in different media (e.g.[42,44]). Such evaluations give confidence in the model's predictions. To ensure that the accuracy of growth-rate predictions generalizes well, we strongly recommend validating growth rates on substrates or in media conditions from which biomass composition and parameters like Growth-Associated Maintenance (GAM) and Non-Growth Associated Maintenance (NGAM) costs were not experimentally derived, as done in [42]. GAM represents the energy expenditure needed to support a certain rate of biomass growth and NGAM represents the energy expenditure required for a cell or organism to survive without any net growth [45]. These values may vary depending on growth conditions, so testing whether the values measured in one set of conditions generalizes to others is important. Otherwise, future users may use a model with, for example, another common media composition and find – or worse yet, simply not notice – that the resulting predictions do not accurately reflect essential characteristics of the organism's actual metabolism.

**Table 1**: The most common model validation strategies in Flux Balance Analysis, what these methods tell us, limitations, and important considerations for researchers and/or readers, and examples of these methods' implementation in the literature.

| Method | Information Content | Limitations | Use case | Examples |
|---|---|---|---|---|
| **Comparison of growth/no-growth on one or more substrates** | Presence/absence of reactions necessary for substrate utilization and biomass synthesis. | Validation is qualitative, only indicating the existence of metabolic routes. Does not test the accuracy of predicted internal flux values | Useful when viability/nonviability of different growth conditions is of interest. Unlike a growth-rate comparison, does not indicate whether the efficiency of biomass synthesis is realistic. | [42,44,59,89] |
| **Comparison of growth-rates on one or more substrates** | Consistency of metabolic network, biomass composition, and maintenance costs with observed efficiency of substrate-to-biomass conversion. | Provides quantitative information on the overall efficiency of substrate conversion to biomass, but is uninformative with respect to accuracy of internal flux predictions. | When done across multiple substrates and conditions, this validation gives confidence in the predicted efficiency with which the model produces biomass. Useful when identifying growth-limiting factors. | [42,59,90] |
| **Comparison of in vivo and in silico knockout lethality** | Presence/absence of biosynthetic reactions necessary for substrate use and growth. | Care is needed to reduce incorrect predictions from many different factors, including optimization method and biomass composition changes in response to knockout. | Critically important to perform when designing growth-coupled knockout strategies [47,49,91]. | [50,90,92] |
| **Comparison of FBA predictions with MFA fluxes** | Accuracy of internal flux predictions. | Few MFA flux maps for most organisms, making this validation impossible or requiring comparison with an MFA flux map taken for a very different experimental conditions. | Important when the intended use of FBA modeling requires that the predictions of specific internal flux values be accurate. | [55,59,93,94] |

A related approach involves comparing growth/no-growth of gene knockout strains to FBA predictions to address whether the metabolic pathways used in the model mirror the biological system. Experimentally verified lethal knockouts that appear nonlethal *in silico* point to alternative routes the model can use to grow (e.g. [46]). Conversely, *in silico* lethality predictions not confirmed by experiment suggest the model is missing isoforms or alternative reaction routes. Researchers sometimes use algorithms to identify knockouts that couple biomass accumulation to flux through a reaction for biotechnological applications [47–49]. This requires that models accurately predict growth/no-growth phenotypes for gene knockouts, but previous work in a model of *Saccharomyces cerevisiae,* for example, shows that FBA performs poorly at predicting the synthetic lethality of double-knockouts, making this a serious concern [50]. When performing such validations, one must keep in mind that imposed constraints and decisions made during the model construction or optimization process may implicitly or explicitly add the predictions one is trying to validate into the model, rendering the exercise meaningless. This makes clear documentation of the assumptions used in the modeling process key for reviewers and readers to assess the epistemic value of the validations that are reported.

It is crucial to note that the methods discussed above do not validate the internal flux predictions made by FBA. In well-characterized systems there may be a wealth of known metabolic functionalities that an organism can carry out and evaluating whether the model can also give some assurance of realistic model behavior. In [51,52], 288 metabolic processes known to take place in mammalian cells were evaluated in models of human and mice models, though it was only the ability to carry out the processes at all, and not the actual flux values, that were evaluated. In favorable cases, individual internal fluxes can be quantitatively estimated *in vivo* using independent methods and compared directly to ones from a predicted flux map to provide a powerful form of validation. For example, in a preprint study from our group [53] the ratio of the cyclic electron flow (CEF) to linear electron flow (LEF) fluxes in photosynthesis predicted by FBA was evaluated against CEF/LEF ratios from fluorescence measurements for validation purposes. Though less specific, the sum of FBA-predicted values for fluxes that produce and/or consume a product (such as $CO_2$) can also be compared to experimental measurements. However, validations of internal flux predictions across the network requires comparing FBA flux maps with high quality ones from 13C-MFA. Unfortunately, such 13C-MFA flux maps are time-consuming to generate, making this "gold-standard" validation rare. Comparison is also complicated by the underdetermined nature of most FBA optimizations, which can result in large feasible ranges for the individual fluxes being compared against the corresponding flux values obtained from 13C-MFA, making the validation less stringent. FBA optimizations that assume parsimony [11,54] tend to yield narrower flux ranges, but this advantage may come at the cost of neglecting other plausible objective functions that might be more accurate. Despite these limitations, some studies, have evaluated the accuracy of FBA against 13C-MFA-estimated flux maps (e.g. [21,55–59]), with mixed results.

To increase the reliability of FBA flux maps, we believe that comparisons against 13C-MFA flux maps should be more widely adopted. This would allow model structure improvement and a thorough analysis of the predictive power of alternative objective functions. For individual biological systems, validation under a limited number of conditions could increase the value of FBA flux maps that can then be predicted for a wide range of conditions and genetic changes. For the wider FBA field, developing useful objective functions (e.g., ones that exploit omic data) also involves FBA/MFA comparisons. Improving confidence in the accuracy of FBA flux maps is valuable because generating validated 13C-MFA flux maps for all systems and conditions of interest is impractical. 13C-MFA requires substantial experimental work for each set of conditions and is unsuitable for many multicellular tissues and organisms where the required combination of extended periods of metabolic steady state, controlled provision of informative, non-perturbing labeled substrates, and obtaining enough labeling data cannot be achieved. This FBA-empowered future for systems biology and biotechnology requires well-validated MFA flux maps, so we turn our attention to validation in MFA.

*Validation in 13C-MFA*

13C-MFA flux estimates are typically validated based on the goodness-of-fit between measured labeling data and the corresponding values generated by the network model after the optimization of model parameters. The goodness-of-fit is represented by the sum of squared residuals (SSR) where each residual is weighted by dividing it by its experimental variance. The $\chi^2$-test of goodness-of-fit, which is built into commonly used 13C-MFA software [60–62], is then used to test whether the SSR falls within the 95% confidence interval expected for the defined number of degrees of freedom (DOF). Since its development as a validation method in 13C-MFA [25], the $\chi^2$-test has been widely used and has been useful in the validation of 13C-MFA metabolic models inferred from genome annotations [63–66].

However, as described in [67] and [68], the use of the $\chi^2$-test can be problematic in 13C-MFA for a number of reasons. When upper- and lower-bounds are imposed on estimated flux parameter values, this makes accurate estimation of the effective DOF for the $\chi^2$-test difficult [68]. It can also be difficult to accurately determine errors in the MID measurements made for 13C-MFA, resulting in distortion of the variance weighted SSR values that are being compared against the 95% Confidence Interval [67].

In addition to these technical difficulties with properly applying the $\chi^2$-test, problems arise from how the test is implemented into the model development process in the course of a typical 13C-MFA study. Especially for eukaryotic systems, 13C-MFA flux modeling generally involves making iterative changes to the model based on how well it can explain the data – as assessed informally and by the $\chi^2$-test – followed by refinement and assessment of the data based on this agreement. For example, if the data do not allow the fluxes between the same metabolite in different compartments to be determined, they may be merged in the model or additional measurements may be made to resolve them. Metabolites may also be excluded from the model due to inconsistency between their simulated vs. measured MIDs causing the model to fail the $\chi^2$-test, sometimes accompanied by biochemical or analytical justifications for their exclusion. The difficulty of accurately quantifying MID measurement errors, mentioned earlier, may be addressed by arbitrarily increasing the assumed measurement error, which reduces the deduced precision of flux estimates to take into account the potential for error sources not accounted for by experimentally observed scatter [67]. This process is a natural consequence of the diversity and uncertainty of the metabolic architecture of different systems and is a valid form of exploratory data analysis and model building. However, altering the model by excluding specific datapoints and adding additional fluxes or metabolites until the $\chi^2$-test passes, and then relying on this very same test as validation is statistically unsafe. As in the case of an FBA model validation in which the predictions being validated against have been implicitly introduced to the model itself, final validation of a 13C-MFA model with the same data used to make it acceptable, as quantified by the $\chi^2$-test, does not constitute a real validation. It also can naturally lead to over- or under-fit models, which we discuss below in the section on model selection.

Due to these difficulties, we propose that the $\chi^2$-test, as it is currently used, should be used as one of multiple lines of evidence to consider when validating a 13C-MFA model. One way to address the issue of using the $\chi^2$-test for both model development and validation is to reserve a portion of the dataset only for final model validation. This practice of holding out a subset of the data to be used exclusively for validation is standard statistical practice [34] in other areas of systems biology and, conveniently, can also be used for model selection [67].

In the absence of direct experimentally measurable fluxes, independent measurements that can be measured or inferred from empirical measurements *in vivo* provide an important ground-truth value to compare with flux estimates and can complement the use of the $\chi^2$-test for validation. An example of this can be found in the plant 13C-MFA literature, where independent measurements of the relative rates of oxygenation and carboxylation by the enzyme RuBisCO can be compared with 13C-MFA flux estimates [69–71] . In [70] for example, our group compared predicted values for the relative rates of oxygenation and carboxylation by the enzyme RuBisCO in photosynthesis versus inferred values from stomatal conductance and other empirical measurements. This led us to conclude that labeling data from whole tissue extracts was insufficient to accurately estimate photorespiratory fluxes without information on the compartmentation of certain metabolites. Despite the strength of this form of validation, it is infrequently practiced.

Another little used but potentially valuable approach to validation is the corroboration of key features of 13C-MFA models using independent modeling methods. In [71], simplified compartmental kinetic models yielded analytical solutions predicting that overall labeling time courses should take the form of sums of exponential rate components. Fitting labeling data to these exponential models and applying statistical model selection techniques provided independent corroboration of the overall architecture of the 13C-MFA model that was used to obtain a detailed flux map.

Returning to goodness-of-fit, one must also keep in mind what information is taken into consideration and the effect of the assumed network architecture. In INST-MFA, where time-course labeling data is used, metabolite pool sizes are both estimable parameters and constrainable modeling inputs. When pool sizes are not provided as empirical measurements, pool size estimates are typically imprecise and inaccurate [72]. The inaccuracy of these estimates is not usually interpreted as an impediment to publishing 13C-MFA results and according to [72], leaving out pool size information does not adversely affect flux estimate accuracy. Flux estimates are not, however, always robust against misspecifications of the network model [67]. Exclusion of pool size information provides greater flexibility in fitting experimental data, allowing robustness against model misspecifications at the expense of not detecting them [72]. We propose that a useful next step for this field would be to routinely measure and include pool size estimates to improve the detection of incorrect model architectures. This introduces the matter of model selection.

**Model Selection in 13C-MFA**

As discussed earlier, model development in 13C-MFA is an iterative process. Alternate models developed during this process may differ in their numbers of reactions and metabolites, resulting in different DOF. Adding model parameters can result in overfitting when these extra DOF lead the 13C-MFA optimization to fit noise rather than biological signal. Model selection techniques can be used to avoid this overfitting and to select the most statistically supported model among alternatives. The development of FBA models can also involve deciding between alternative architectures. However, comparison and selection of such models from sets of alternatives based on their predictions' deviations from empirical measurements is uncommon, so we focus our attention on 13C-MFA.

Model misspecification can result in missing important fluxes, incorrectly estimating the rates of modeled fluxes, or incorrectly estimating the precision of flux estimates. In a study our group performed of central metabolic fluxes in the oilseed crop *Camelina sativa* [71], previously published model architectures that passed the $\chi^2$-test of goodness-of-fit [70] were nonetheless shown to be missing an important set of metabolic reactions involving the movement of carbohydrates to and from the vacuole. In [67], *in silico* examples of sub-optimal model selection resulting in flux estimates that fall outside of the 95% confidence intervals for those same fluxes generated using the correct model architecture are provided, showing the potential for biased flux estimates when model selection is not properly performed. Finally, the literature on "Genome-scale-13C-MFA" has provided evidence that the exclusion of many reactions peripheral to the metabolic network under consideration (typically core metabolism) in 13C-MFA can result in artificially narrow confidence intervals. Genome-scale-13C-MFA involves estimating a flux map by minimizing deviation between predicted and measured isotopic labeling, but using the kind of genome-scale metabolic network more typically used for FBA analyses [73,74]. In studies on the cyanobacterium *Synechococcus elongatus* [75,76], it has been shown that the substantially larger genome-scale 13C-MFA models achieved better fits to the labeling data, that these reductions in SSR were statistically justified, and that the original models of core metabolism underestimated the uncertainty in a number of flux estimates by ignoring alternative metabolic pathways that could also explain patterns in the labeling data [74]. The examples above demonstrate that rather than being a statistical curiosity, model selection (or the lack thereof) can have serious implications for the accuracy and reliability of flux modeling results.

A number of approaches to model-selection can be found in the 13C-MFA literature, with different approaches being taken in different studies. The simplest is selecting the model with the smallest SSR. This method does not work when the DOF of the compared models are different, as increasing the DOF in a model inevitably allows it to fit a given data set better. This may be accounted for in an informal way by noting the change in DOF (e.g. [71]), or in a more statistically rigorous way using the extra-sum-of-squares test [77,78] or information criteria [79,80]. The most common model selection approach used in 13C-MFA is an informal method using the $\chi^2$-test, wherein models are iteratively modified until a model and dataset pass the test, or where a selection of alternative models are evaluated and the one that passes the test by the widest margin is selected [67,81–83]. These approaches have been used, for example, to demonstrate that the isotopic labeling data of co-culture systems cannot be adequately described by modeling with a single-culture 13C-MFA model [84,85], to provide evidence for the operation of previously undescribed fluxes in mammalian cells [86], and to detect missing reactions in metabolic network reconstructions from genome annotations or that are needed to describe the metabolism of mutant *E. coli* strains [57,63].

However, the previously mentioned limitations of the $\chi^2$-test for model validation also affect its usefulness for model selection and models failing the test due to these limitations can lead to the addition of statistically unjustified metabolites or reactions to the model until it passes [67]. We refer to the $\chi^2$-test-based methods as "informal" model selection because when multiple models are evaluated, they are not directly or formally compared to determine whether the additional parameters in more complex models are statistically justified, which can naturally lead to the selection of overfit models.

The general approach of avoiding overfitting by evaluating models based on their performance on a set of data not used during the fitting process is widely used in statistics (e.g. cross-validation techniques [87]). The validation-based approach taken in [67] brings this best-practice separating fitting and testing data sets to avoid the pitfalls discussed above and, in our view, represents a substantial advancement in model selection in 13C-MFA. This method divides the labeling dataset into training and validation subsets and then estimates fluxes in alternative models using the training data. These alternative models' flux maps, and their accompanying predicted MIDs, are then compared on the basis of their agreement with the validation MID data. The model whose flux map results in the smallest SSR when compared with this validation data is selected. The authors generated synthetic labeling data from a predefined "correct" model and assessed the ability of their new method and other model selection techniques to identify this correct model from a set of alternatives. The validation-based approach accomplishes this more consistently than existing model selection methods, including $\chi^2$-test-based methods, and does so irrespective of the value of the measurement error in the labeling datasets. The incorrect models selected by other methods contain flux estimates that fall outside the 95% confidence intervals of the fluxes from the correct model, highlighting the importance of model selection for obtaining accurate flux estimates [67]. The generation of MID data in additional labeling experiments to precisely measure all fluxes in a network [26–33] provide the reserved validation datasets needed for [67]. This means that for 13C-MFA studies that already require a parallel labeling approach, implementation of this more rigorous model selection approach is simply a matter of setting aside a subset of data to evaluate alternative model architectures.

This approach can be extended in INST-MFA by using metabolite pool size measurements in the selection process. Individual pool sizes are sensitive to the local kinetic parameters and will fit poorly when reaction networks are incompletely specified [72]. We therefore suggest that validation-based model selection using pool size measurements as input measurements is a promising prospective model selection approach for INST MFA (**Figure 2**). As the authors of [67] note, the optimal model selected by their method should be subjected to a final validation to assess model quality. A model architecture may be selected by the model selection process but result in a substantial deviation of some metric from independently measured values. For this final validation, a combination of the $\chi^2$-test, independent experimental measurements, and alternative modeling approaches can be used. Keeping in mind both the trade-off between goodness-of-fit and model complexity and the multiple ways in which 13C-MFA model predictions can be validated will ensure that flux estimates are as accurate and robust as possible.

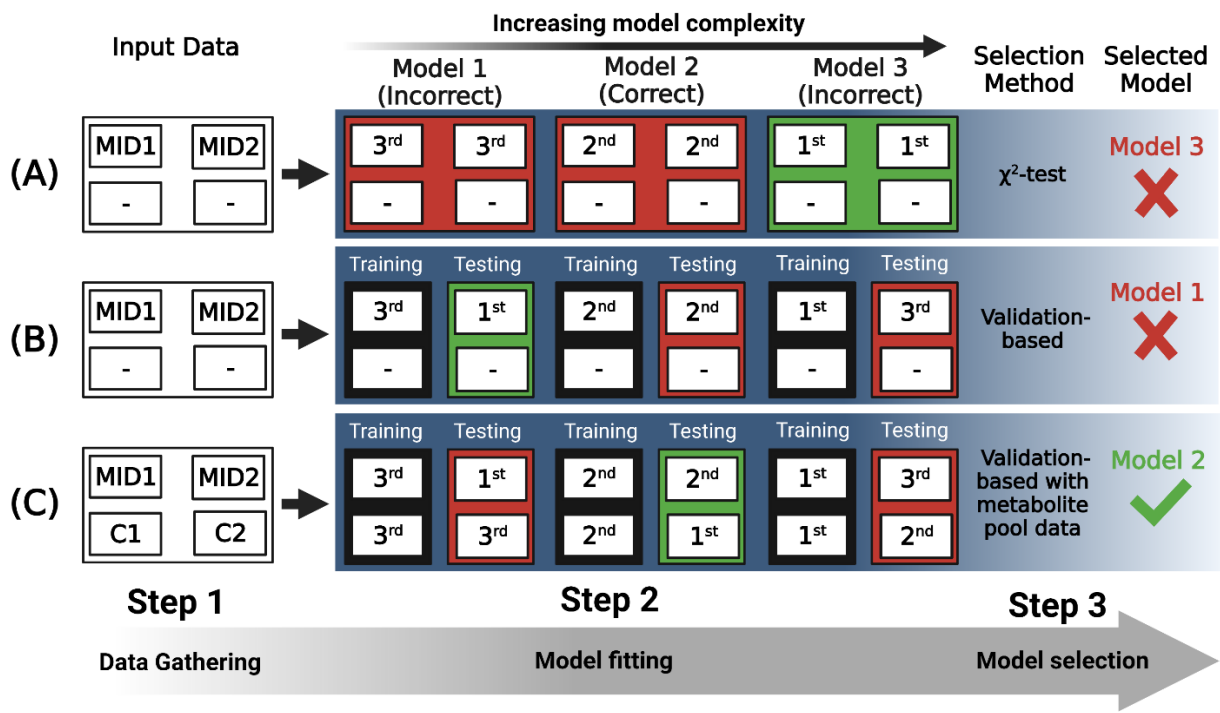

**Figure 2:** *Approaches to model selection for 13C-MFA. Metabolic network models 1-3 having increasing complexity are compared. Model 2 in this example is the correct description of the network. **(A)** Labeling data (MID1 & MID2) are gathered and, for each model, agreement between model output and these data is optimized. The $\chi^2$-test of goodness-of-fit is used to assess each model fit and these model fits are ranked 1st, 2nd, or 3rd, with the 1st passing the test by the widest margin and being selected as the most statistically well-supported model. **(B)** Labeling data are split into "training" and "testing" subsets and agreement between model output and the "training" data is optimized. The Sum-of-Squared Residuals (SSR) is then calculated for each model from the deviation between its output and the "testing" data. The model fits are then ranked 1st, 2nd, and 3rd, with the 1st having the lowest SSR and being selected. **(C)** Labeling data and metabolite pool data (C1 and C2) are gathered and split into "training" and "testing" subsets. For each model, agreement between model output and these data is optimized. The Sum-of-Squared Residuals (SSR) is then calculated for each model from the deviation between its output and the "testing" data. The model fits are then ranked 1st, 2nd, and 3rd, with the 1st having the lowest SSR and being selected. The inclusion of metabolite pool size data into both the "fitting" and "testing" datasets provides more data to go off of when evaluating goodness-of-fit, potentially increasing the likelihood of identifying the correct model from a set of alternatives.*

### Future Directions

We believe that validation and selection deserve greater attention from the flux analysis community and suggest that implementing the approaches highlighted in this perspective will improve the accuracy and reliability of constraints-based metabolic modeling and flux estimates.

However, we also recognize that some approaches suggested here, such as the use of pool size measurements, can be extremely difficult to implement in practice. Smith *et al.,* in a recent publication on isotopically non-stationary MFA of *Arabidopsis thaliana* heterotrophic cell culture metabolism, highlighted that although they realized that pool size data could potentially be used to improve the accuracy and precision of their flux predictions, the experimental difficulty of measuring the concentrations of metabolites distributed across multiple subcellular compartments made this prohibitively difficult [88]. As in all areas of science, then, the development of consensus best-practices in evaluation of and inference from data and models must arise at the intersection of rigorous statistical theory and experimental practicalities. However, we believe that researchers engaged in constraint-based metabolic modeling as well as readers of modeling studies benefit when the limitations of present validation and selection practices are clarified.

Several matters call for investigation before definitive recommendations can be made on best-practice. At present, it is not clear how to appropriately weight the contributions to flux estimation of unambiguous direct flux measurements like substrate uptake, which typically have relatively large standard deviations, against MIDs, which frequently have much smaller standard deviations but whose relationship to fluxes depends on model structure and whose measured values may be offset by unknown analytical effects. Likewise, it is unclear how best to deal with those not infrequent MID measurements that have extremely small, but imprecisely measured, standard deviations, which can exert too much control over the fitting process.

Finally, we would like to conclude by emphasizing that the process of careful validation and model selection can lead to the generation of models that are not only more quantitatively sound, but that yield exciting scientific insights (e.g. [71,85,86]).

**Key Points**

- A number of approaches exist for the validation of Flux Balance Analysis model predictions, but most validate only one or a small number of features of the model. When drawing conclusions from or using another research group's Flux Balance Analysis model, the type and extent of validations performed on that model must be taken into account to ensure its suitability.

- The $\chi^2$-test of goodness-of-fit is widely used for model validation in 13C-Metabolic Flux Analysis and is an important tool, but technical limitations and issues with its incorporation into iterative model-building processes make it important to corroborate the $\chi^2$-test with other forms of validation.
- Advancements in the use of parallel labeling experiments have set the stage for more sophisticated model selection in 13C-MFA. Comparison of models on the basis of their prediction accuracy when evaluated against an out-of-sample dataset can help researchers select the best-supported model from a selection of alternatives.

## Acknowledgements

This research was supported by the Office of Science (BER), U.S. Department of Energy, Grant no DE-SC0018269 (J.A.M.K., Y.S-H.). This work is supported, in part, by the NSF Research Traineeship Program (Grant DGE-1828149) to J.A.M.K. This publication was also made possible by a predoctoral training award to J.A.M.K. from Grant T32-GM110523 from National Institute of General Medical Sciences (NIGMS) of the NIH. Its contents are solely the responsibility of the authors and do not necessarily represent the official views of the NIGMS or NIH. Figures made using BioRender.com.

## Author Contributions

J.A.M.K and Y.S-H conceptualized the manuscript. J.A.M.K. wrote the manuscript with input and editing from Y.S-H.